\documentclass[10pt,a4paper, twocolumn,superscriptaddress]{revtex4-1}
\usepackage{amsmath}
\usepackage{amsfonts}
\usepackage{braket}
\usepackage{amssymb}
\usepackage{graphicx}
\usepackage{color}
\usepackage{textcomp}
\usepackage{soul}
\usepackage{comment}

\newcommand{\ga}{\gamma}

\newcommand{\de}{\delta}
\newcommand{\ep}{\varepsilon}

\begin{document}

\title{Nondeterministic quantum computation via ground state cooling and ultra-fast Grover's algorithm}









\author{P. V. Pyshkin}
\affiliation{Beijing Computational Science Research Center, Beijing 100084, China}
\affiliation{Department of Theoretical Physics and History of Science, The Basque Country University (EHU/UPV), PO Box 644, 48080 Bilbao, Spain}
\affiliation{Ikerbasque, Basque Foundation for Science, 48011 Bilbao, Spain}

\author{Da-Wei Luo}
\affiliation{Beijing Computational Science Research Center, Beijing 100084, China}
\affiliation{Department of Theoretical Physics and History of Science, The Basque Country University (EHU/UPV), PO Box 644, 48080 Bilbao, Spain}
\affiliation{Ikerbasque, Basque Foundation for Science, 48011 Bilbao, Spain}

\author{J. Q. You}
\affiliation{Beijing Computational Science Research Center, Beijing 100084, China}

\author{Lian-Ao Wu}
\thanks{To whom all correspondence should be addressed. lianao.wu@ehu.es}
\affiliation{Department of Theoretical Physics and History of Science, The Basque Country University (EHU/UPV), PO Box 644, 48080 Bilbao, Spain}
\affiliation{Ikerbasque, Basque Foundation for Science, 48011 Bilbao, Spain}
\date{\today}

\begin{abstract}

Over the last decades, there have been many proposals for quantum computation. One of the promising candidates is adiabatic quantum computation (AQC). 
 The central idea of AQC is about finding the ground state of a system with a problem Hamiltonian via particular adiabatic passages, starting from an initialized ground state of a simple Hamiltonian. One disadvantage of AQC is the significant growth of necessary runtime, 
 in particular when there are quantum phase transitions during the AQC passages. 
 Here we propose a nondeterministic ground state cooling quantum computation model based on selective projection measurements on an ancilla coupled to the system with the problem Hamiltonian previously cooled by conventional techniques. We illustrate the model by Grover search problem and show that our nondeterministic model requires a constant or at most logarithmic runtime and can also get rid of possible difficulties arising from preparation of the ground state of the simple Hamiltonian. 

\end{abstract}

\maketitle


\bigskip

Building a quantum computer is an intriguing task for scientists and engineers. The reason behind this interest follows from the theoretical possibility of solving some problems much faster on quantum computers than on their classical counterparts. The well-known paradigms are the prime factorization algorithm~\cite{Shor1994} and the search algorithm through unsorted databases~\cite{Grover}. To this end, different approaches of quantum computation have been proposed. A widely popular example is quantum circuit model proposed by Deutsch in~1989~\cite{Deutsch1989}. In this protocol, universal quantum computation is performed by the sequential application of unitary operations as quantum gates to a single qubit or groups of qubits. The initial state of qubits is assumed to be a product state, and after the computation process, the final state of these qubits contains the solution of the target problem. In contrast, a one-way quantum computation proposed in Ref.~\cite{one_way} is a completely different approach. 
It consists of consecutive single-qubit projective measurements on an initially-prepared highly entangled cluster state. The target problem in the one-way computation is encoded in a particular type of entangled states. A third model is adiabatic quantum computation~\cite{Adiabatic_qc_1} based on the adiabatic theorem~\cite{Born1928}. The solution of the target problem is encoded into the ground state~$\ket{\psi_0}$ of a problem Hamiltonian~$H_0$. The core task of AQC is to physically achieve and then read out the ground state. However, generally, since it is NP-hard~\cite{GS-NP}, finding ground state of a large system cannot be done analytically or numerically on classical computers. 
The AQC scheme employs a time-dependent Hamiltonian $H(t) = a(t)H_1 + b(t)H_0$, where $H_1$ is an auxiliary Hamiltonian with an easily achievable ground state $\ket{\psi_1}$. Both $a(t)$ and $b(t)$ are slowly varying time-dependent functions with $a(0)=1, a(T)=0, b(0)=0$, and $b(T)=1$, where $T$ is the runtime of the computational process. The adiabatic theorem states that at any instant the system follows the original stationary and yet time-dependent eigenstate of the instantaneous $H (t )$ if parameters $a(t)$ and $b(t)$ satisfy the adiabatic condition. In this way the eigenstate ~$\ket{\psi(0)}=\ket{\psi_1}$ of the initial Hamiltonian $H(0)$ evolves into the eigenstate $\ket{\psi(T)}=\ket{\psi_0}$ of the final $H(T)$, which encodes the solution to the target problem and will be read out. Importantly, it has been shown that AQC is \emph{universal} and equivalent to the universal quantum circuit model~\cite{aqcuniversal}.

Adiabatic quantum computation has two disadvantages. One is the preparation of the initial ground state of the auxiliary Hamiltonian~$H_1(0)$, which could be difficult for systems such as nuclear spins in NMR. If the initial state is mixed, after adiabatic passage the final output state will remain mixed, such that the results may be invalid. The other is the runtime~$T$ of the adiabatic evolution, which increases dramatically with the database size or the problem complexity, in particular when there are quantum phase transitions. To overcome these disadvantages, here we propose a new universal quantum computation model: \textit{Ground state cooling} quantum computation~(GSCQC). It is axiomatic that the ground state cooling of the problem Hamiltonian $H_0=H(T)$ leads to the same ground state $\ket{\psi(T)}=\ket{\psi_0}$ as in adiabatic quantum computation. Therefore, GSCQC is a universal quantum computation model, the same as AQC. In contrast with AQC, our GSCQC does not have the second disadvantage, because we cool the system governed by the problem Hamiltonian directly. Thus GSCQC and AQC appear to confront the same challenge -- ground state cooling, however GSCQC aims at directly cooling down to the solution to the target problem while AQC to the ground state of the auxiliary Hamiltonian.  

Over the years, there have been many proposals for ground state cooling of specific systems such as atoms and ions. Now we propose to use a generic {\em nondeterministic} ground state cooling technique and exemplify, by Grover's search problem, that cooling to the ground state of a problem Hamiltonian can be realized by a set of selective measurements of a specifically designed ancilla system. In addition, we should emphasize that unlike the proven square-root speedup in Grover's deterministic algorithm, our ultrafast algorithm for Grover's problem is not deterministic and should have its own restrictions. 





\section*{Results}
\subsection*{Hamiltonian design}
To solve a specific problem, our GSCQC protocol starts with designing a problem Hamiltonian whose ground state corresponds to the solution, as in AQC. Different from the AQC, we then need to design an ancillary system such that we can realize ground state cooling with high probability. 

Grover's search problem is modelled by considering a set $\{ \ket{0}, \ket{1}, \dots , \ket{N-1}\}] $ of $N$ orthogonal states in a finite Hilbert space. One of these states, $\ket{w}$, is the solution to the search problem, and our task is to find it. For any given state $\ket{j}$, the system as an Oracle gives a yes/no answer to the question ``is it true that $\ket{j} = \ket{w}$?''. The Oracle Hamiltonian reads~\cite{Adiabatic_qc_2},
\begin{equation}\label{H_0}
H_0 = \ep \left( \vphantom{1^1} \mathbb{I} - |w \rangle\langle w| \right),
\end{equation}
where $\ep$ is the strength of Hamiltonian, and $\mathbb{I}$ is the identity matrix. From~(\ref{H_0}) it follows that $H_0\ket{j}=\lambda \ket{j}$, where $\lambda = 0$ for $\ket{j}=\ket{w}$ and $\lambda = \ep$ for the rest $\ket{j}\neq\ket{w}$.

Because $\ket{w}$ is the non-degenerate ground state, the system described by Hamiltonian~(\ref{H_0}) will reach this state after the ground state cooling process. Importantly, we have to ensure ourselves that the system has to be in its ground state {\em before} reading it out. This requirement determines the strategy of our proposed GSCQC protocol, where cooling process is divided into two steps. First, the system is cooled by conventional techniques to the lowest achievable temperature $T$, and afterwards we apply our shot cooling scheme~\cite{Li2011}, which has been verified experimentally in Ref. \cite{Xu2014}.

Assume that after the conventional cooling, our system is in a Gibbs thermal state described by a density matrix:
\begin{align}\label{initial_rho}
&\rho_{or}(0) = p_0\ket{w}\bra{w} + p_1\sum_{n\neq w}\ket{n}\bra{n} , \nonumber \\
&p_0 = \dfrac{1}{1 + (N-1)e^{-\ep/kT}}, \, p_0 + (N-1)p_1 = 1,
\end{align}
where $k$ is the Boltzmann constant, $p_0$ is the probability of the system being in the ground state, $p_1$~is the probability of finding system in one of excitation states $|n \rangle$ $(n \neq w)$. 
We show that by using our GSCQC protocol, the initial thermal state can be driven to a pure (or almost pure) solution state: $\rho_{or}(0)\xrightarrow{GSCQC} \ket{w}\bra{w}$.

\begin{figure}[t]
 	\begin{center}
 		\includegraphics{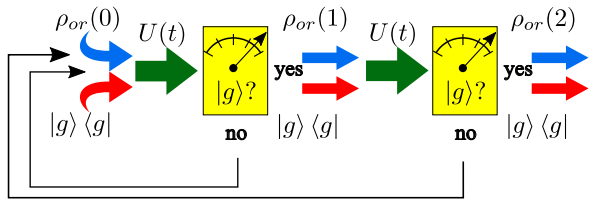}
 	\end{center}
 	\caption{\textbf{Sketch of our GSCQC protocol.} The initial product states of the system~$\rho_{or}(0)$ and ancilla qubit~$\ket{g}\bra{g}$ become correlated due to the joint evolution~$U(t)$. After projection measurement~$\ket{g}\bra{g}$ on ancilla the next step of cooling process is applied if the outcome is~$\ket{g}$, and the state of the system is changed to~$\rho_{or}(1)$. Otherwise, the process has to run again from the beginning. }
 	\label{fig1}
 \end{figure}

It has been shown theoretically~\cite{Li2011} and experimentally~\cite{Xu2014} that the system ground state cooling can be achieved by a sequence of joint unitary evolutions and selective measurements on an ancilla coupled to the system. In our case we select a qubit as the ancilla. The total Hamiltonian of the system (or Oracle) and ancilla reads
\begin{align}\label{H_total}
H =& H_0 + \ga \left(\ket{g}\bra{g}-\ket{e}\bra{e}\right) + \nonumber \\
 &\de\sum_{n=0}^{N}\left( \ket{n,e}\bra{n+1,g}+h.c.  \right),
\end{align}
where $\ket{g}$ $(\ket{e})$ is the ground (excitation) state of the ancilla qubit, $\gamma$ corresponds to the energy splitting of the ancilla qubit, and $\delta$ is the coupling strength between the ancilla and Oracle where we consider a periodic boundary condition: $\ket{N}\equiv \ket{0}$ and set $\ep=1$ and~$\hbar=1$ throughout the paper. Importantly, the second and third terms in Hamiltonian~(\ref{H_total}) do not contain any information about the unknown answer~$\ket{w}$. The initial density matrix of the whole system is $\rho(0) = \rho_{or}(0)\otimes\ket{g}\bra{g}$. In {\bf{Methods}}, we show how the system is cooled down to the state $\rho_{or}(M)$ at the~$M$-th measurement on ancilla~\cite{Li2011,Hiromichi}. The shot cooling is carried out in $M (\ge 1)$ iterations. In each iteration, we start with a different density matrix, which evolves under the Hamiltonian~(\ref{H_total}) over a particular period of time~$t$. After the evolution, one makes a projective selective measurement~$\ket{g}\bra{g}$ on the ancilla. After measurement, the density matrix becomes new. If the outcome of measurement is~$\ket{g}$ then we repeat and start the next iteration, otherwise the system will be reset, as in Fig.~\ref{fig1}. Furthermore we show that by properly choosing time~$t$ and parameters~$\ga$ and $\de$, one can achieve ground state cooling with a very high probability after few ancilla measurements. In what follows, we will introduce two different strategies for implementation of our GSCQC protocol. The first is specifically designed for Grover's search problem and the second is for finding the solution to a general problem, illustrated likewise by Grover's problem. 

\subsection*{Algorithm of the ground state cooling}
Under the standard basis,  the Hamiltonian and its propagator matrices can be represented by the direct sum of two-dimensional submatrices or blocks, and these blocks can be classified into three types only. One of these blocks (denoted as $0$) corresponds to the state~$\ket{w}$  and another (denoted as~$1$) to the~$\ket{w+1}$ state. The third type corresponds to all other $N-2$ states (denoted as~$2$, see Eq.(\ref{h_blocks}) in {\bf Methods}). The first strategy works by engineering the Hamiltonian in such a way that the propagator for the second type block is a swap operator $\begin{pmatrix}
0 & 1 \\
1 &  0 \\
\end{pmatrix}$. To achieve this we set~$\gamma = 0$ and $\delta_{1}t = \pi/2 + \pi j$, where $j=0,1,2,\dots$. After evolution and a measurement on the ancilla, if outcome is $\ket{g}$, the Oracle state becomes $\rho_{or}(1) = (p_0\ket{w}\bra{w}+p_1\ket{w+1}\bra{w+1})/(p_0+p_1)$. In the second step, we engineer a swap operator for first block by setting~$\gamma = -1/2$ and again $\delta_{2} t = \pi/2 + \pi j$, $j=0,1,2,\dots$. Given that the outcome of the ancilla measurement is $\ket{g}$, we can conclude with certainty that the Oracle state is $\rho_{or}(2) = \ket{w}\bra{w}$. The probability~$p_s$ of achieving successful ground cooling therefore is
\begin{multline}\label{probabil_2_steps}
p_s = p_0\prod_{i=1,2}\left(\cos^2\left[\sqrt{\delta_i^2+1/4}\frac{\pi/2 + \pi j}{\delta_i}\right] + \right.\\
\left.\frac{1/4}{\delta_i^2+1/4}\sin^2\left[\sqrt{\delta_i^2+1/4}\frac{\pi/2 + \pi j}{\delta_i}\right] \right) \leq p_0
\end{multline}
where $i=1,2$ is the step number. The probability~$p_s$ can be made approximately equal to the initial probability $p_0$ by choosing~$\delta_i\rightarrow 0$ ($t_i \rightarrow \infty$). Equation~(\ref{probabil_2_steps}) concludes that~$p_s\geq p_0(1-\delta_m^2/(\delta_m^2+1/4))$ where $\delta_m=\max\{\delta_1;\delta_2\}$. Assume that one has~$K$ copies of the same Oracles, the probability~$p(K)$ of achieving the above strategy in at least one copy of Oracles is~$p(K)=1-(1-p_s)^K\approx1-(1-p_{0})^K$. For instance, if~$p(K)=0.99$ is required,  one has $K\geq7$ for $p_0=1/2$, and $K\geq44$ for~$p_0=1/10$.

The second strategy assumes that there is no flexibility to adjust parameters in the Hamiltonian during cooling process. The goal of this process is that the  0-type $2\times2$ evolution matrix becomes diagonal and the 2-type matrices come to be as close to a swap operator as possible after $M$ measurements on the ancilla, given that all outcomes are $\ket{g}$. 
When~$t=2\pi$, the optimal parameters for the search algorithm are $\ga\approx0.059$ and $\de\approx0.236$ 
(see {\bf Methods} for details).

 \begin{figure}[t]
	\begin{center}
		\includegraphics{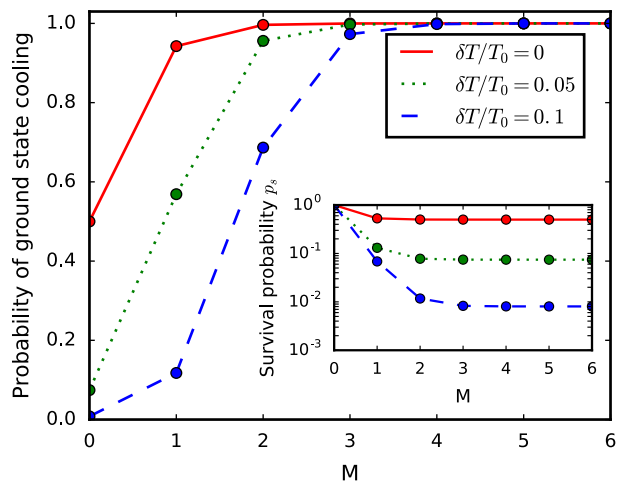}
	\end{center}
	\caption{\textbf{Probabilities as a function of $M$.} The probability of
finding ancilla in its ground state under different temperatures. Here the database size is $N=10^{23}$ and $T = T_0 + \delta T$, where $T_0 = \ep (k\ln N)^{-1}$. Inset: Survival probability as a function of $M$ at different temperatures.}
	\label{fig2}
\end{figure}

\subsection*{Initial conventional cooling as a quantum computation tool}

The effectiveness of GSCQC dramatically depends on the initial temperature~$T$. The characteristic temperature~$ T_0$, which we define by ~$p_0 = 1/2$ in Eq.~(\ref{initial_rho}), is $T_0 = \dfrac{\ep}{k\ln N}$. It has been shown~\cite{grover_number} that the minimum size ($N$) of oracle database from which quantum Grover's search becomes more advantageous than the classical counterpart is $N\approx10^{22}$. The energy gap~$\ep$ in the experimental NMR realization of the Grover search algorithm~\cite{nmr_adiabatic} is~$\ep\approx 8.3\times10^{-19}$~erg which corresponds to the resonance frequency~$125.76$~MHz of~$^{13}$C in a magnetic field of $11.2$~Tesla. Using these data we estimate~$T_0\approx1.2\times10^{-4}$~K. This temperature is very low, but it is achievable by using modern cooling techniques~\cite{low-temp-book, cooling_BEC_500pk}. 
Proposal of quantum computation based on electron spin resonance in solids~\cite{ESR_QC} allows to have $\exp(-\ep/kT)\approx0.02$ by using a combination of strong pre-polarization fields and laser pulses at cryogenic temperature $T = 4.2$~K. This corresponds to $T_0\approx 0.3$~K for the Hamiltonian~(\ref{H_0}) with $N=10^{23}$. These examples imply that we can expect experimental realizations of the ground state even with conventional cooling,  reading out the correct {\em answer} with high probability. We thus claim that for the problem Hamiltonian, the conventional ground state cooling already acts as a non-deterministic quantum computer, a conventional cooling induced quantum effect like superconductivity. This seems to be the simplest version of quantum computing. After reading out, the state should be finally checked and confirmed by substituting it into eigenequation of the problem Hamiltonian. Now the issue arises with the final check and confirmation. Unlike explicitly-written problem Hamiltonians, our problem Hamiltonian is a {\em black-box or unknown} Hamiltonian hidden in the Oracle, with which we are not able to check whether or not the state satisfies the eigenequation of the unknown Hamiltonian. In this sense, conventional ground state cooling cannot complete the full task of quantum computation, and we therefore need the second step. 

\subsection*{Efficiency of ground state cooling}

Equation~(\ref{initial_rho}) shows that small temperature variations around $T_0$ can cause dramatic changes to the probability $p_0$. As discussed earlier, our second step uses shot cooling to avoid errors from temperature fluctuations, obtain a near~$\approx100\%$ probability of achieving the right answer, and more importantly provide a complete readout scheme for unknown Hamiltonians.

Let us consider~$N=10^{23}$ and denote the temperature of the system after the conventional cooling as $T = T_0 + \delta T$, where we consider worse cases when $\delta T\geq 0$. Figure~\ref{fig2} shows the probability of achieving the answer state~$\ket{w}$ after $M$ ancilla measurements, given that all outcomes are $\ket{g}$, for different values of~$\delta T/T_0$. After a few such ancilla measurements, the probability approaches ~$1$. The inset of Fig.~\ref{fig2} shows the survival probability~$p_s$ versus $M$, which approaches a constant $p_0$. For instance, if  $T = T_0$ and ~$p_0 = 1/2$ initially, after $M=3$ ancilla measurements  it is shown that the probability to be in ground state~$\ket{w}$ is about~$0.999$ while $p_s\approx 1/2$.

We now consider the effect of the temperature fluctuations.   For a given probability $P_{cooling}$,  different temperature fluctuations requires different ancilla measurement times $M$. 
For our first strategy, $P_{cooling}=1$ for two ancilla measurements. Fig.~\ref{fig2} shows for the second strategy, where $M$ increases with $\delta T/T_0$. 
Assume that maximum value of $\delta T/T_0$,  the minimal $M_p$ ancilla measurements are required, i.e.
\begin{align}\label{log_law_3}
&M_p < \log_{1/b_2}\dfrac{P_{cooling}N}{1-P_{cooling}-aP_{cooling}}, \nonumber\\
&a = \left( \frac{1}{N} \right)^{\dfrac{1}{1 + \delta T/T_0}},
\end{align}
with the probability $P_{cooling} < 1 - N^{-1/(1+\delta T/T_0)}$. 

\begin{figure}[t]
	\begin{center}
		\includegraphics{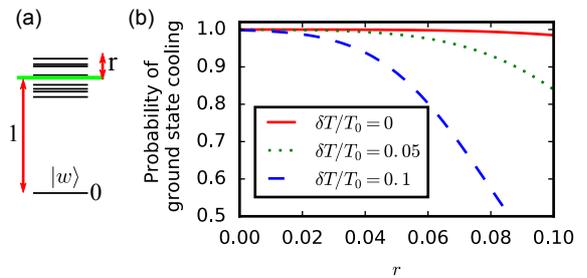}
	\end{center}
	\caption{\textbf{Probabilities as a function of the gap $r$. } (a) Schematic illustration of Oracle's energy levels. (b) Minimal probability  of finding ancilla in its initial state with $M=4$ as a function of the gaps $2r$ between low and high excitation states under different temperature. The database size is $N=10^{23}$ and $T = T_0 + \delta T$, where $T_0 = \ep (k\ln N)^{-1}$.} \label{fig4}
\end{figure}

\section*{Discussion}

We have illustrated our nondeterministic GSCQC model by Grover's search problem.  The model has two steps to find the ground state of a problem Hamiltonian. We first notice that the conventional cooling itself may act as quantum computation to find the ground state for a {known} problem Hamiltonian. For an {unknown} problem Hamiltonian, our second step finds the ground state by ancilla measurements, which is conceptually proposed by ref.~\cite{Li2011}.  We design two strategies for ancilla measurements:  $M=2$ and does not depend on the database size $N$ for the first strategy designed specifically for Grover's search problem and in the second strategy $M$ depends logarithmically on $N$ but for a general problem Hamiltonian, as exemplified by Fig.\ref{fig4}. 

It is interesting to note that if the ground state~$\ket{w}$ is known, the second step of the GSCQC model can be further simplified to one ancilla measurement, by using two Hadamard gates acting on ancilla qubit, one control-$U$ gate and one ancilla measurement, where $U$ is the propagator of the black-box Oracle Hamiltonian~(\ref{H_0}), $U = \exp(-i\pi H_0)$. However, the simplified scheme cannot be applied to an unknown~$\ket{w}$, because realizing control-$U$ gate with black-box~$H_0$ Hamiltonian is impossible as proven in~\cite{unknown-u}. Intuitively, the control-$U$ gate may be physically given by the Hamiltonian $H_{c-U} = H_0 - \sum_{n\neq w}\ket{n,g}\bra{n,g}$, which implies a contradiction that one has to have pre-knowledge about the known ground state~$\ket{w}$.

A direct generalization is the case when excitation states are not fully degenerate. Figure~\ref{fig4}~(a) assumes that the energy gaps~$2r$ between low and high excitation states are smaller than the gaps between the ground state and the first excitation state. In~Fig.\ref{fig4}~(b) we show the cooling probability of the Oracle after $M=4$ ancilla measurements as a function of the gap~$r$, indicating that small energy gaps between low and high excitation states do not change the cooling possibility.


\section*{Methods}

After $M$ ancilla measurements, given that all outcomes are $\ket {g}$, the density matrix of the  system is~\cite{Hiromichi}:
\begin{equation}\label{rho_transf_total}
\begin{aligned}
\rho_{or}(M) = \dfrac{V^M\rho_{or}(0)V^{\dag M}}{\mathrm{Tr}(V^M\rho(0)V^{\dag M})}, \\
\end{aligned}
\end{equation}
where $V = \braket{g|e^{-iHt}|g}$. Due to the block-diagonal structure of $H$,  it is easier to treat each $2\times2$ submatrices of $H$ and $\rho$ separately. 
It can be checked that the Hamiltonian~(\ref{H_total}) is block-diagonal with three types of $2\times2$ submatrices or blocks,
\begin{align}\label{h_blocks}
&h_0 = \begin{pmatrix}
1-\ga & \de \\
\de & \ga \\
\end{pmatrix}, \nonumber \\
&h_1 = \begin{pmatrix}
-\ga & \de \\
\de & 1+\ga \\
\end{pmatrix}, \nonumber\\
&h_2 = \begin{pmatrix}
1-\ga & \de \\
\de &  1 + \ga \\
\end{pmatrix}.
\end{align}
There are $N-2$ blocks of $h_2$ types and one block of $h_0$ and $h_1$ each in the Hamiltonian. The block $h_0$  corresponds to the solution $\ket{w}$. The density matrix of the system stays block-diagonal throughout the shot cooling process, and we denote the corresponding $2\times2$ blocks of the total density matrix as $\rho_0(M), \rho_1(M)$, and $\rho_2(M)$ respectively. The corresponding initial blocks are  $\rho_0(0)=p_0\mathrm{diag}(0,1)$, and $\rho_1(0)=\rho_2(0)=p_1\mathrm{diag}(0,1)$. 
 The evolution of the density matrix blocks between two consecutive measurements is governed by the corresponding blocks of Hamiltonian. After $M$ ancilla measurements, the blocks of the density matrix becomes
\begin{align}\label{rho_steps}
&\rho_0(M) = W_0(M)\mathrm{diag}(0,1), \nonumber\\
&\rho_1(M) = W_1(M)\mathrm{diag}(0,1), \nonumber\\
&\rho_2(M) = \frac{W_2(M)}{N-2}\mathrm{diag}(0,1)
\end{align}
with
\begin{align}\label{W_b_def}
&W_0(M)+W_1(M)+W_2(M)=1, \quad W_0(0)=p_0, \nonumber\\
&W_1(0)=(N-2)^{-1}W_2(0)=p_1, \nonumber\\
&W_i(M+1) = b_iA(M)W_i(M), \; i=0,1,2.\nonumber\\
&A(M) = \left( \sum_i b_i W_i(M) \right)^{-1}, \nonumber\\
&\begin{pmatrix}
1-b_i & . \\
. & b_i \\
\end{pmatrix} = e^{-ih_it}\begin{pmatrix}
0 & 0 \\
0 & 1 \\
\end{pmatrix}e^{ih_it},
\end{align}
where $b_i$ corresponds to the $\ket{g}$ state of the ancilla and the dots in a matrix in Eq.~(\ref{W_b_def}) are arbitrary numbers irrelevant to the shot cooling procedure. $W_0(M)$ is the probability of the Oracle system being in the answer state~$\ket{w}$. From Eq.~(\ref{W_b_def}) we can obtain
\begin{equation}\label{Wi-from-W0}
W_i(M) = \frac{W_i(0)b_i^M}{\sum_{j}W_j(0)b_j^M}.
\end{equation}
Eq.~\eqref{Wi-from-W0} is used for numerical simulation in {Fig.~\ref{fig2}. Following~\cite{gsc_paper} we require simultaneously: (i) $b_0=1$ (dictated by the answer block) and (ii) $b_2\rightarrow0$ (corresponding to the second type block). 
Using Eqs.~(\ref{h_blocks}) and~(\ref{W_b_def}), the requirement (i) is
\begin{equation}\label{first_condition}
\dfrac{\de^2}{\de^2 + (1/2 - \ga)^2}\sin^2\sqrt{\de^2 + (1/2 - \ga)^2}t = 0.
\end{equation}
By setting $t=2\pi$ we get from Eq.~(\ref{first_condition})
\begin{equation}\label{ep-de-relation}
\de^2 = \ga(1-\ga) + \frac{1}{4}\left( n^2 -1  \right), \; n=1,2,3\dots.
\end{equation}
We take~$n=1$. The requirement (ii) means that the following expression should have a maximum value:
\begin{equation}\label{second_condition}
\dfrac{\de^2}{\de^2 + \ga^2}\sin^2\sqrt{\de^2 + \ga^2}t.
\end{equation}
The last requirement allows us to estimate $\delta$ and $\gamma$ for any given $t$. 

\section*{Acknowledgments}
We acknowledge grant support from the Basque Government (Grant No. IT986-16), the Spanish MICINN (Project No. FIS2015- 69983-P), and the Basque Country University UFI (Project No. 11/55-01-2013), the National Key Research and Development Program of China (No. 2016YFA0301200) and the NSAF (Nos. U1330201 and U1530401).

\bibliographystyle{ieeetr}

\end{document}